\documentclass[conference]{IEEEtran}
\IEEEoverridecommandlockouts
\usepackage{cite}
\usepackage{amsmath,amssymb,amsfonts}
\usepackage{graphicx}
\usepackage{textcomp}
\usepackage{xcolor}
\def\BibTeX{{\rm B\kern-.05em{\sc i\kern-.025em b}\kern-.08em
    T\kern-.1667em\lower.7ex\hbox{E}\kern-.125emX}}

\usepackage{cite}
\usepackage{graphicx,color,epsfig,rotating}
\usepackage{amsfonts,amsmath,amssymb,bbm}
\usepackage{algorithm}
\usepackage{algpseudocode}
\usepackage{subfigure}
\usepackage{amsmath}
\usepackage{cite}
\usepackage{mdwtab}
\usepackage{subfigure}
\usepackage{placeins}
\usepackage{psfrag, graphicx}
\usepackage[latin1]{inputenc}
\usepackage{amssymb}
\usepackage{makeidx}
\usepackage{epstopdf}
\usepackage{enumitem}

\long\def\comment#1{}

\newfont{\bbb}{msbm10 scaled 700}

\newfont{\bb}{msbm10 scaled 1100}

\newcommand{\bv}{{\bf b}}

\newcommand{\yv}{{\bf y}}
\newcommand{\zv}{{\bf z}}

\newcommand{\Xm}{{\bf X}}

\newcommand{\be}{\begin{equation}}
\newcommand{\ee}{\end{equation}}
\newcommand{\bea}{\begin{eqnarray}}
\newcommand{\eea}{\end{eqnarray}}

\newcommand{\RED}{\color[rgb]{1.00,0.10,0.10}}
\newcommand{\BLUE}{\color[rgb]{0,0,0.90}}

\newcommand{\cs}[1]{\texttt{\symbol{`t}#1}}

\newtheorem{defn}{Definition}
\newtheorem{theorem}{Theorem}
\newtheorem{remark}{Remark}

\begin{document}



\title{A Practical Algorithm Design and Evaluation for Heterogeneous Elastic Computing with Stragglers}

\author{Nicholas Woolsey$^{1}$, J\"org Kliewer$^{2}$, Rong-Rong Chen$^{1}$ and Mingyue Ji$^{1}$
\thanks{The authors are with the Department of Electrical Engineering,
University of Utah, Salt Lake City, UT 84112, USA. (e-mail: nicholas.woolsey@utah.edu, rchen@ece.utah.edu and mingyue.ji@utah.edu)}
}

\author{
    \IEEEauthorblockN{ Nicholas Woolsey$^{1}$, J\"org Kliewer$^{2}$,
		Rong-Rong Chen$^{1}$ and Mingyue Ji$^{1}$ }
	\IEEEauthorblockA{$^1$University of Utah, \;\; $^2$New Jersey Institute of Technology\\
		Email: \{nicholas.woolsey@utah.edu,
        jkliewer@njit.edu,
		 rchen@ece.utah.edu,
		mingyue.ji@utah.edu\}}

}

\maketitle

\thispagestyle{empty}
\pagestyle{empty}


\begin{abstract}
Our extensive real measurements over Amazon EC2 show that the virtual instances  often have different computing speeds even if they share the same configurations. This motivates us to study  heterogeneous Coded Storage Elastic Computing (CSEC) systems where machines, with different computing speeds, join and leave the network arbitrarily over different computing steps. In CSEC systems, a Maximum Distance Separable (MDS) code is used for coded storage  such that the file placement does not have to be re-defined with each elastic event. Computation assignment algorithms are used to minimize the computation time given computation speeds of different machines. While previous studies of heterogeneous CSEC do not include stragglers -- the slow machines during the computation, we develop a new framework in heterogeneous CSEC that introduces straggler tolerance. Based on this framework, we design a novel algorithm using our previously proposed approach for heterogeneous CSEC such that the system can handle any subset of stragglers of a specified size while minimizing the computation time.  Furthermore, we establish a trade-off in computation time and straggler tolerance. Another major limitation of existing CSEC designs is the lack of practical evaluations using real applications. In this paper, we evaluate the performance of our designs on Amazon EC2 for applications of the power iteration and linear regression.  Evaluation results show that the proposed heterogeneous CSEC algorithms 
outperform the state-of-the-art designs by more than $30\%$.
\end{abstract}



\section{Introduction}
\label{section: intro}
Coded Storage Elastic Computing (CSEC) was introduced by Yang et al. \cite{yang2018coded} as an effective method to perform distributed computing on elastic cloud systems where machines have limited storage capacity. Here, elasticity means that machines can join and leave the network periodically as machines may be reserved or freed from other higher priority jobs. The main advantage of CSEC is that  machines store coded data such that a computation on the data can be recovered as long as enough machines are available. This is in contrast to storing uncoded data where if some machine leaves with a particular subset of  data, then a computation on the entire data set may not be recovered; or, in order to tolerate the same number of preempted machines, the required storage/computation overhead is larger. The use of coded storage alleviates
the need to re-distribute the storage with each elastic event which could greatly slow down the cloud computing system. 

While coded computing has been  studied 
in the literature for straggler mitigation  \cite{lee2017speeding,dutta2016short,ferdinand2016anytime,yu2020straggler,kliewer2019coded,karakus2017straggler,halbawi2018improving,suh2017matrix,maity2018robust,Aktas2018straggler,wang2018fundamental,ye2018communication,wan2020distributed}, 
CSEC is different from these works because, at the beginning of a computation step,
we know which machines are available  
{\it a priori}. 
This allows us to design an effective computation assignment to the 
machines. Assume we use a Maximum Distance Separable (MDS) code to encode the data matrix with a recovery threshold of $L$ and each machine stores one of  coded sub-matrices (cs-matrices). If there are exactly $L$ available machines for the given time step, then each machine will perform computations on all locally available data and the master machine will collect all computation results to recover the original computation task. However, if there are more than $L$ machines, we can strategically assign computations such that each machine performs computations on a fraction of its local coded data, effectively reducing the overall computation time. 
The overall computation is split into many steps, where in each step we know which machines are available, even though their availability may change over computation steps. The original CSEC design \cite{yang2018coded} studied  homogeneous CSEC systems where machines have the same computation speed. It proposed a cyclic homogeneous computation assignment design  to minimize the computation load at each machine such that no computation redundancy is present in the network. 

Based on our  measurements of the computation speeds on Amazon EC2 instances, we  found that these virtual instances often have different computation speeds even if they share  the same configuration. Hence, in \cite{woolsey2020heterogeneous},
 for  heterogeneous elastic computing systems where machines have different computation speeds,
we reformulated the CSEC problem as a combinatorial optimization problem with the objective of minimizing the overall computation time, instead of per machine computation load as in \cite{yang2018coded}.
This combinatorial optimization problem was solved using a novel approach, which decomposed this problem into two sub-optimization problems consisting of a convex optimization problem and a combinatorial ``filling'' problem, where an optimal solution was found.  
Then, in \cite{dau2020optimizing}, Dau et al. introduced a new concept called Transition Waste (TW) that measures the difference between the total number of changes and the necessary changes of the computation assignment. The authors proposed novel algorithms to find the minimum TW for some parameter settings in  homogeneous CSEC systems. 

However, a question often arises for CSEC designs. {\it What happens if an elastic event occurs during a computation step?} Computation assignments are designed such that the set of available machines is known during a computation step. If a machine does not respond during a computation step, it
is deemed as a failed node
and we have to re-assign computations to recover the intended computation. In addition, a machine may not respond for reasons other than an elastic event such as becoming unpredictably slow, unresponsive or otherwise simply fails. In general, we will label any one of these events during a computation step as a {\it straggler}. 
Adding straggler tolerance in CSEC is an interesting problem that has not been studied in the literature.
Given 
the MDS coded storage placement used in \cite{yang2018coded}, 
Kiani et al. in \cite{kiani2021CEC} proposed hierarchical computation assignment algorithms 
for homogeneous CSEC systems. However, the performance evaluation in \cite{kiani2021CEC} were done using computer simulations, where stragglers are modeled to take place with certain 
probabilities, instead of using practical cloud computing systems such as Amazon EC2 or Microsoft Azure. In addition, the computation task is only matrix-matrix multiplication rather than real  applications. 


In this paper, our focus is to 1) incorporate  straggler toleration into the heterogeneous CSEC framework and 2) conduct actual performance evaluations 
over Amazon EC2 without any assumptions or models of stragglers. 
In particular, using the existing MDS coded storage design,  we will show that straggler tolerance can be incorporated into our proposed heterogeneous CSEC optimization framework \cite{woolsey2020heterogeneous}. The main advantage of our proposed approach is to provide straggler tolerance by adjusting  computation assignments. 
Furthermore, we show that there is an interesting trade-off between the computation time of heterogeneous CSEC system and the number of stragglers that can be tolerated. One extreme case is the heterogeneous CSEC with no straggler tolerance and the other extreme case is to maximize straggler tolerance using the coded computing design of \cite{lee2017speeding}. We will present a general framework that includes these designs and show there is a straggler tolerance and computation time trade-off. 

Our contributions are summarized as follows:
\begin{enumerate}
  \item We develop a new framework in heterogeneous CSEC that introduces straggler tolerance. Based on this framework, we design a novel algorithm using our previously proposed approach for heterogeneous CSEC such that the system can handle any subset of stragglers of a specified size and the computation time is minimized.
  \item 
  The proposed new heterogeneous CSEC framework introduces a trade-off between computation time and straggler tolerance of CSEC systems. An analytical expression is derived to characterize the optimized computation time as a function of the computation speeds and the straggler tolerance. 
  \item We validate our proposed CSEC designs by actual implementations on Amazon EC2 instances for applications of both the  power iteration algorithm  
  and 
  linear regression.  
  In particular, the computation assignment for the proposed CSEC is based on real-time measurements of the computation speeds of Amazon EC2 instances.
\end{enumerate}

\paragraph*{Notation Convention}
We use $|\cdot|$ to represent the cardinality of a set or the length of a vector
and $[n] := [1,2,\ldots,n]$. 

\section{Network Model and Problem Formulation}
\label{sec: Network Model and Problem Formulation}

A set of $N$ machines each store a coded sub-matrix derived from a $q\times r$ data matrix, $\boldsymbol{X}$. The coded matrices are defined by an $N\times L$ generator matrix $\boldsymbol{G}=[g_{n,\ell}]$ which represents an MDS code such that any $L$ rows of $\boldsymbol{G}$ are invertible. The data matrix, $\boldsymbol{X}$, is row-wise split into $L$ disjoint, $\frac{q}{L}\times r$ sub-matrices, $\boldsymbol{X}_1,\ldots ,\boldsymbol{X}_L$. Each machine $n\in[N]$ stores the coded sub-matrix (cs-matrix)
\be
\boldsymbol{\tilde{X}}_n = \sum_{\ell=1}^{L}g_{n,\ell}\boldsymbol{X}_\ell.
\ee
of size $\frac{q}{L} \times r$.

The machines collectively perform matrix-vector computations over multiple computation steps. In a given step only a subset of the $N$ machines are available to perform matrix computations. More specifically, in computation step $t$, a set of available machines $\mathcal{N}_t \subseteq [N]$ aims to compute
\be
\boldsymbol{y}_t = \boldsymbol{X}\boldsymbol{w}_t
\ee
where $\boldsymbol{w}_t$ is some vector of length $r$. The machines of $[N]\setminus \mathcal{N}_t$ are preempted. 

The machines of $\mathcal{N}_t$ do not compute $\boldsymbol{y}_t$ directly. Instead, each machine $n\in \mathcal{N}_t$ computes the set
$\mathcal{V}_{n} = \left\{ v = \boldsymbol{\tilde{X}}_n^{(i)}\boldsymbol{w}_t : i \in \mathcal{W}_{n} \right\}$, 
where $\boldsymbol{\tilde{X}}_n^{(i)}$ is the $i$-th row of $\boldsymbol{\tilde{X}}_n$ and $\mathcal{W}_{n}\subseteq\left[ \frac{q}{L}\right]$ is the set of rows assigned to machine $n$ in time step $t$. 
\begin{defn} {\bf (Computation load)} 
The computation load vector, $\boldsymbol{\mu}$, is defined as 
\be \label{eq: compload_vector}
\mu[n] = \frac{|\mathcal{W}_{n}|}{
\left( \frac{q}{L} \right)}, \;\; \forall n \in \mathcal{N}_t, 
\ee
which is the fraction of rows of the corresponding stored cs-matrix computed by machine $n$ in time step $t$. \hfill $\Diamond$ 
\end{defn}
Note that, $\boldsymbol{\mu}$, $\mathcal{V}_{n}$ and $\mathcal{W}_{n}$ change with each computation step, but reference to $t$ is omitted for ease of disposition. Moreover, the machines have varying computation speeds defined by the strictly positive vector, $\boldsymbol{s}$, which is known for each computation step and defined as follows.

\begin{defn}
{\bf (Computation Speed)} The computation speed vector $\boldsymbol{s}$ is a length-$N$ vector with elements $s[n]$, $n\in[N]$, where $s[n]$ is the speed of machine $n$ measured as the inverse of the time it takes machine $n$ to compute all rows of its cs-matrix. 
The normalized computation speed is the computation speed divided by the average computation speed among all available machines. 
\hfill $\Diamond$
\end{defn}

The computation time is dictated by the machine that takes the most time to perform its assigned computations, defined as follows. 

\begin{defn} {\bf (Computation Time)}
The computation time in a particular computation step is defined as 
\be \label{eq: comptime}
c(\boldsymbol{\mu}) = \max_{n\in \mathcal{N}_t} \frac{\mu[n]}{s[n]}. 
\ee
\hfill $\Diamond$
\end{defn}
In a given computation step, for each $i\in\left[ \frac{q}{L}\right]$, at least $L$ machines perform the vector-vector multiplication with the $i$-th row of their local cs-matrix and $\boldsymbol{w}_t$. The results are sent to a master machine which can resolve the elements of $\boldsymbol{y}_t$ by the MDS code design.

A computation assignment is defined by $F$ disjoint sets of rows, $\boldsymbol{\mathcal{M}}_t  = (\mathcal{M}_{1},\ldots,\mathcal{M}_{F} )$ whose union is $\left[ \frac{q}{L}\right]$. Then, $F$ sets of machines, $\boldsymbol{\mathcal{P}}_t = (\mathcal{P}_{1} , \ldots , \mathcal{P}_{F} )$, are defined such that $\mathcal{P}_f \subseteq\mathcal{N}_t$ and $|\mathcal{P}_f|=L$ for all $f\in [F]$. The rows of $\mathcal{M}_{f}$ are assigned to the machines of $\mathcal{P}_{f}$. In other words, the rows computed by machine $n\in \mathcal{N}_t$ in time step $t$ are in the set
$\mathcal{W}_{n} = \bigcup \left\{ \mathcal{M}_f : f\in[F],  n \in \mathcal{P}_{f}\right\}$ 
and therefore, $\boldsymbol{\mu}$ is a function of $\left(\boldsymbol{\mathcal{M}}_t, \boldsymbol{\mathcal{N}}_t \right)$. The sets $\mathcal{M}_{1},\ldots,\mathcal{M}_{F}$ and $\mathcal{P}_{1} , \ldots , \mathcal{P}_{F}$ and $F$ may vary with each computation step based on  machines' availability.

In a given computation step $t$, our goal is to design the task assignments, $\boldsymbol{\mathcal{M}}_t$ and $\boldsymbol{\mathcal{N}}_t $, such that the computation $\boldsymbol{y}_t = \boldsymbol{X}\boldsymbol{w}_t$ can be recovered even when some machines are stragglers that do not provide their assigned computations to the master machine. We define $S$ as the straggler tolerance of the network, such that the computation can be recovered even when $S$ of the available machines become stragglers. Furthermore, let $\mathcal{S}_t\subset\mathcal{N}_t$ be the subset of the available machines that become stragglers. When designing the computation assignment, $\mathcal{S}_t$ is not known and in general  can be any subset of 
$S$ available machines.

Then, we aim to design the computation assignment that minimizes the computation time of (\ref{eq: comptime}) resulting from the computation load vector defined in (\ref{eq: compload_vector}).
In computation step $t$, given $\mathcal{N}_t$ and $\boldsymbol{s}$, the optimal computation time, $c^*$, is the minimum of computation times defined by all possible task assignments,  $\left(\boldsymbol{\mathcal{M}}_t, \boldsymbol{\mathcal{N}}_t \right)$, such that $S$ stragglers can be tolerated and the computation can be recovered. In particular $c^*$ is the optimal value of the following combinatorial optimization problem. 
\begin{subequations} 
\label{eq: opt}
\begin{align}
&\min_{\left(\boldsymbol{\mathcal{M}}_t, \boldsymbol{\mathcal{N}}_t \right)}c\left(\boldsymbol{\mu}\left(\boldsymbol{\mathcal{M}}_t, \boldsymbol{\mathcal{N}}_t \right)\right) \\
&\text{s.t. }\bigcup_{\mathcal{M}_f \in \boldsymbol{\mathcal{M}}_t}\mathcal{M}_f=\left[ \frac{q}{L}\right],\label{eq: optprob_assign}\\
&\;\;\;\;\;\; |\mathcal{P}_f\setminus \mathcal{S}_t | \geq L \;\; \forall \mathcal{P}_f \in \boldsymbol{\mathcal{P}}_t, \;\; \forall \mathcal{S}_t \subset \mathcal{N}_t, \;\; |\mathcal{S}_t|=S\\
&\;\;\;\;\;\; |\boldsymbol{\mathcal{M}}_t|=|\boldsymbol{\mathcal{P}}_t|. 
\end{align}
\end{subequations}

\begin{remark}
  The main difference between this problem formulation and that of \cite{woolsey2020heterogeneous} is the consideration of the straggler machine set $\mathcal{S}_t$. The computation assignment is designed such that the computation on each row set is guaranteed to be computed at least $L$ times across the machines (excluding stragglers). In general, all possibilities of $\mathcal{S}_t$ must be considered which could be any subset of 
$S$ available machines.
\end{remark}

\section{Examples} 
\label{sec: example}
We first compare the CSEC designs \cite{yang2018coded,woolsey2020heterogeneous} to the straggler tolerant design of \cite{lee2017speeding}. Then, we present examples of the proposed design which bridge the gap between the two design methodologies.  Homogeneous systems are discussed first, followed by  heterogeneous systems where machines have varying computation speeds.

\subsection{Homogeneous Examples}

Consider a set of $N_t=5$ available machines in a computation step. Each machine $n$ has the same computation speed of $s[n]=1$. The storage at these machines adopts a $(5,3)$-MDS code where the machines store the coded matrices of $\boldsymbol{\tilde{X}}_1=\boldsymbol{X}_1$, $\boldsymbol{\tilde{X}}_2=\boldsymbol{X}_2$, $\boldsymbol{\tilde{X}}_3=\boldsymbol{X}_3$, $\boldsymbol{\tilde{X}}_4=\boldsymbol{X}_1+\boldsymbol{X}_2+\boldsymbol{X}_3$, and $\boldsymbol{\tilde{X}}_5=\boldsymbol{X}_1+2\boldsymbol{X}_2+4\boldsymbol{X}_3$. Here $L=3$, is the recovery threshold of the MDS code. With these $5$ available machines, we can assign computations such that each row is computed by at least $L$ machines.

{\it Homogeneous Cyclic CSEC Design \cite{yang2018coded}, $(S=0)$}: Using the cyclic homogeneous design, the rows of the coded matrices are split into $F=5$ disjoint subsets. The first row set is computed at machines $1$ through $3$, the next row set is computed at machines $2$ through $4$ and so on. Following the notation of the problem formulation, the row sets are $\mathcal{M}_i=\{1+(i-1)\frac{q}{5L},\ldots, i\frac{q}{5L} \}$ for $i\in[5]$ and the machine sets are $\mathcal{P}_1=\{1,2,3 \}$, $\mathcal{P}_2=\{2,3,4 \}$, $\mathcal{P}_3=\{3,4,5 \}$, $\mathcal{P}_4=\{4,5,1 \}$, and $\mathcal{P}_5=\{5,1,2 \}$.  The main advantage of this design is that each machine $n$ computes $\mu[n] = \frac{3}{5}$ of its local cs-matrix. Since each machine has a speed of $1$, we  see that the computation time is $c=\frac{3}{5}$. However, this design is susceptible to stragglers. Since each row set is assigned to exactly $L$ machines, if any one machine becomes a straggler, then the matrix-vector multiplication cannot be recovered.

{\it Straggler Design of \cite{lee2017speeding}, $(S=2)$}: Alternatively, each machine computes the entire local cs-matrix. In other words, the assignment is $F=1$, $\mathcal{M}_1=\{1,\ldots , \frac{q}{L} \}$, $\mathcal{P}_1=\{1,2,3,4,5 \}$. In this case, it is clear that $S=2$ stragglers can be tolerated, since we only need computations recovered from any $L=3$ machines. However, this is at a cost of computation time which increases to $c=1$, or the time for a machine to compute all rows of its local cs-matrix.

{\it A Homogeneous CSEC Straggler Design, $(S=1)$}: The question remains if the computation time can be reduced when the straggler tolerance is only $S=1$ straggler. It turns out there is a computation-straggler tolerance trade-off if we use another computation assignment. For example, using the principles of the cyclic assignment, we can assign computations to $L+S=4$ machines instead of just $L$ machines. That is, we define $F=5$ row sets, $\mathcal{M}_i=\{1+(i-1)\frac{q}{5L},\ldots, i\frac{q}{5L} \}$ for $i\in[5]$, and assign these to the machine sets of $\mathcal{P}_1=\{1,2,3,4 \}$, $\mathcal{P}_2=\{2,3,4,5 \}$, $\mathcal{P}_3=\{3,4,5,1 \}$, $\mathcal{P}_4=\{4,5,1,2 \}$, and $\mathcal{P}_5=\{5,1,2,3 \}$, respectively. Note that each row set is assigned to $4$ machines, such that if any one machine becomes a straggler, the computation on that row set can be recovered. Also, note that the computation time is now $c=\frac{4}{5}$ since each machine performs computations on $4$ row sets each containing $\frac{1}{5}$ of the rows.

\subsection{Heterogeneous Examples}

Assume that there are $N_t=5$ available machines which have the same storage as in the homogeneous examples. Again, $L=3$ and each row computation required computations from $L=3$ machines to be recovered. However, this time, the machines have heterogeneous computation speeds defined by 
$\boldsymbol{s} = [1,1,2,2,3]$ 
such that machine $5$ is the fastest  with a computation speed of $3$ and machines $1$ and $2$ are the slowest  with a computation speed of $1$.

{\it Heterogeneous Algorithm CSEC Design \cite{woolsey2020heterogeneous}, $(S=0)$}: We define a computation assignment that minimizes the computation time. It is ideal to assign each machine an amount of computations relative to the speed of the machine. In this case, we define the computation load of the machines to be
$\boldsymbol{\mu}=\left[\frac{1}{3}, \frac{1}{3},\frac{2}{3},\frac{2}{3},1\right]$,
where, for example, machines $1$ and $3$ will compute a $\frac{1}{3}$ and $\frac{2}{3}$ fraction, respectively, of their local cs-matrix. Moreover, the computation load vector sums to $L=3$ such that it allows  each row to be computed $3$ times. A computation assignment that yields this computation load is $F=3$ row sets each with $\frac{1}{3}$ of the rows assigned to the machines of $\mathcal{P}_1=\{1,4,5 \}$, $\mathcal{P}_2=\{2,3,5 \}$ and $\mathcal{P}_3=\{3,4,5 \}$. Each machine takes the same time to perform their assigned computations and the computation time of the network is $c=\frac{1}{3}$. However, there is no straggler tolerance.

{\it Straggler Design of \cite{lee2017speeding}, $(S=2)$}: Each machine computes the entirety of its local cs-matrix. In this case, $S=2$ stragglers are tolerated. However, the computation time is dictated by the slowest machines which are machines $1$ and $2$. Since we plan for the possibility of any machine straggling, we may need to wait for machines $1$ and $2$ to finish their computations. This means the computation time is $c=1$.

{\it Proposed Heterogeneous CSEC Straggler Design, $(S=1)$}: The computation time can be improved if we reduce the straggler tolerance to $S=1$. We assign computations to $L+S=4$ machines yielding a computation load vector of
$\boldsymbol{\mu}=\left[\frac{1}{2}, \frac{1}{2},1,1,1\right]$,  
where machines $3$ through $5$ have a computation load of $1$ and compute the entirety of their local matrix. Adapting the assignment algorithm of \cite{woolsey2020heterogeneous}, the following computation assignment yields this computation load. There are $F=2$ row sets each containing half of the rows assigned to the machines of $\mathcal{P}_1=\{1,3,4,5 \}$ and $\mathcal{P}_2=\{2,3,4,5 \}$, respectively. Notice that each row set is assigned to $4$ machines, such that if any one machine straggles, the result can be recovered. Furthermore, the computation time is defined by the local computation time of machines $1$ through $4$ which each take the same time to finish. The computation time is $c=\frac{1}{2}$.

\section{Proposed Heterogeneous CSEC Design}

Our proposed straggler tolerant CSEC design, given by Algorithm~\ref{algorithm:2}, is obtained by solving the combinatorial optimization (\ref{eq: opt}) in a similar fashion as in \cite{woolsey2020heterogeneous} (line 6 in Algorithm~\ref{algorithm:2}), and measuring (line 14 in Algorithm~\ref{algorithm:2}) and updating (line 4 in Algorithm~\ref{algorithm:2}) the speed vector at each iteration. This algorithm adapts previous CSEC designs \cite{woolsey2020heterogeneous} to assign computations to $L+S$ machines instead of just $L$ machines. We will explain the proposed heterogeneous CSEC design in the following. For completeness, we start with the homogeneous CSEC design first.

{\it Homogeneous CSEC Straggler Tolerant Design}: Given $N_t$ available machines, $\mathcal{N}_t=\{1,2,\ldots ,N_t \}$, define a computation assignment with $F=N_t$ row sets. There are $N_t$ disjoint equally-sized row sets that collectively span all rows: $\mathcal{M}_i=\{1+(i-1)\frac{q}{LN_t},\ldots, i\frac{q}{LN_t} \}$ for $i\in[N_t]$. Then, define a cyclic assignment such that machine set $\mathcal{P}_i=\{i\%N_t,\ldots,(i+L+S-1)\%N_t\}$ for $i\in[N_t]$, where we define $a\%N_t \triangleq a-\left \lfloor \frac{a-1}{N_t} \right\rfloor N_t $ to facilitate the cyclic design.

{\it Proposed Heterogeneous CSEC Straggler Tolerant Design}: Assume there are $N_t$ available machines, $\mathcal{N}_t=\{1,2,\ldots ,N_t \}$ with computation speeds in ascending order $s[1] \leq s[2] \leq \cdots \leq s[N_t]$. Define the computation load vector to be
\begin{equation}
\mu^*[n]=\begin{cases}
c^* s[n] & \text{if } 1\le n \le k^*\\
1 & \text{if } k^*+1 \le n \le N_t,
\end{cases}
\label{eq:optimal_form_hm}
\end{equation}
where $k^*$ is the largest integer in $[N_t]$ such that
\be
\frac{1}{s[k^*+1]} < c^* = \frac{L+S-N_t+k^*}{\sum_{n=1}^{k^*}s[n]} \leq \frac{1}{s[k^*]}.
\label{eq: c_bounds-fork_hm}
\ee
Given the computation load vector $\mu^*$, 
we can obtain the computation 
assignment by applying the assignment algorithm of \cite{woolsey2020heterogeneous} to assign computations to $L+S$ machines instead of $L$ machines (line 6 in Algorithm~\ref{algorithm:2}). This naturally includes straggler tolerance and yields the following theorem.

\begin{theorem}\label{th: CSECstraggler}
  The computation times of the straggler tolerant CSEC designs are:

  {\it Homogeneous design}:
  \be
  c^* = \frac{L+S}{N_t}\cdot\frac{1}{\min_n s[n]}
  \ee

  {\it Heterogeneous design}:
  \be
  c^*=\frac{L+S-N_t+k^*}{\sum_{n=1}^{k^*}s[n]}
  \ee
  where $k^*$ is defined in the heterogeneous design. \hfill $\Box$
\end{theorem}


\begin{remark}
  For both designs, we observe that the computation time $c$ increases with the straggler tolerance, $S$. This demonstrates a trade-off between the computation time and straggler tolerance of the system.
\end{remark}



\begin{algorithm} 
  \caption{Adaptive Straggler Tolerant Coded Storage Elastic Computing}
  \label{algorithm:2}
  \begin{algorithmic}[1]
  \item[ {\bf Input}: $\hat{\boldsymbol{s}}$, $\gamma$, $S$, $T$, $\boldsymbol{w}_1$ ] 
   \hspace*{4cm} 
   \State $\boldsymbol{\nu}\leftarrow \hat{\boldsymbol{s}}$
  \For {$t \in [T]$}
   \State {\bf At Master Machine}:
  \State \quad $\hat{\boldsymbol{s}}\leftarrow \gamma \boldsymbol{\nu} + (1-\gamma)\hat{\boldsymbol{s}}$  { (update estimate of speed vector)}.
  \State \quad $\mathcal{N}_t \leftarrow$ list of available machines 
  \State \quad $(F$, $\mathcal{M}_{1}, \ldots , \mathcal{M}_{F},$ $\mathcal{P}_{1}, \ldots , \mathcal{P}_{F})\leftarrow$ Results of computation assignment algorithm with straggler tolerance of $S$ for available machines $\mathcal{N}_t$ with speeds of $\hat{\boldsymbol{s}}$
  \State \quad Send $\boldsymbol{w}_t$ and $(F$, $\mathcal{M}_{1}, \ldots , \mathcal{M}_{F}$, $\mathcal{P}_{1}, \ldots , \mathcal{P}_{F})$ to Worker Machines
   \State {\bf At Worker Machines}:
   \State \quad $n\leftarrow$ index of worker machine
   \State \quad $\mu \leftarrow$ computation load of worker machine $n$
  \State \quad $\tau_1\leftarrow$ current time
  \State \quad Perform assigned computations based on $(F$, $\mathcal{M}_{1}, \ldots , \mathcal{M}_{F}$, $\mathcal{P}_{1}, \ldots , \mathcal{P}_{F})$
  \State \quad $\tau_2\leftarrow$ current time
  \State \quad $\nu[n] \leftarrow \mu[n]/(\tau_2-\tau_1)$ {(calculate speed based on current computation step)}
  \State \quad Send computations and $\nu[n]$ to Master Machine
  \State {\bf At Master Machine}:
  {after receiving results from $L$ workers}.
  \State \quad $\boldsymbol{w}_{t+1}\leftarrow$ Decode and combine worker results
  \EndFor
  \item[ {\bf Output}: $\boldsymbol{w}_T$ ]
  \end{algorithmic}
\end{algorithm}


\section{Evaluation on Amazon EC2}

\subsection{Evaluation Setup}

We evaluate the proposed algorithm using two representative real applications including the power iteration and linear regression algorithms on Amazon EC2 instances as follows.

{\it Power Iteration}: 
The power iteration algorithm computes the largest eigenvalue and the corresponding eigenvector of a large matrix $\Xm$. In particular, it starts with a vector $\bv_0$, which may be an approximation to the dominant eigenvector or a random vector. The method is described by the recursive relation, $\bv_{k+1} = \frac{\Xm \bv_k}{\|\Xm \bv_k\|}$. The sequence $\bv_k$ converges to an eigenvector associated with the dominant eigenvalue. It can be seen that at each iteration, we can directly apply the proposed Algorithm~1. In particular, 
a dense $60,000$-by-$60,000$ symmetric matrix is row-wise split into $L=10$ sub-matrices which are used to define the coded matrices stored at each machine. 
A vector of length $60,000$ is updated by performing a matrix-vector multiplication in a distributed manner on the available worker machines. The master machine combines the results and normalizes the vector. This process is repeated such that the vector converges to the eigenvector associated with the largest eigenvalue.

{\it Linear Regression}: A linear regression problem is to solve $\min_\bv \|\Xm \bv - \yv\|^2$. One typical approach to solve such algorithm is to use gradient descent $\bv_{k+1} = \bv_k - \eta \Xm^{\rm T} \left(\Xm \bv_k - y\right)$. Letting $\Xm \bv_k - y = \zv_k$,  we see that in each iteration, the algorithm consist of 2 matrix-vector multiplications, which are $\Xm \bv_k$ and $\Xm^{T} \zv_k$. We apply Algorithm~1 to both of them. 
In particular, we use a $200,000$-by-$5,000$ matrix representing a dataset with $200,000$ entries each with $5,000$ features. We follow the distributed gradient descent design of \cite{lee2017speeding,yang2018coded}. Each worker machine stores two coded data matrices, one coded row-wise and the other coded column-wise and $L=10$. These are used to perform matrix multiplication with the data matrix and its transpose  in a distributed fashion to update the weights. 

The network has one \verb"t2.x2large" master machine with $8$ vCPUs and $32$ GiB of memory. The worker machines consist of $10$ \verb"t2.large" instances, each with $2$ vCPUs and $8$ GiB of memory, and $10$ \verb"t2.xlarge" instances, each with $4$ vCPUs and $16$ GiB of memory. Each machine is either a {\it stable machine} or an {\it elastic machine} such that with each iteration, the $12$ stable machines ($6$ \verb"t2.large" and $6$ \verb"t2.xlarge") are always available for computation assignment and the elastic machines ($4$ \verb"t2.large" and $4$ \verb"t2.xlarge") are each available with a probability of $0.5$. The availability of the elastic machines are independent across machines and also each computation step. Note that, any available machine that is assigned computations in a computation step could become a straggler (slowest machine). 
In this case, the master machine will not wait for the $2$ slowest machines (stragglers) to finish their computations (``with straggler" scenario in Figs.~\ref{fig: Power} and \ref{fig: LinRegr}). We define 
the {\em uncoded} and {\em no straggler} scenarios as follows. 




{\it Uncoded}: 
No straggler tolerance is considered. The stable machines include $5$ \verb"t2.large" and $5$ \verb"t2.xlarge" machines which store uncoded matrices of the dataset without redundancy. 
We do not consider elastic machines in this case. 

{\it No Stragglers}: 
No straggler tolerance is considered. The stable machines consist of $5$ \verb"t2.large" and $5$ \verb"t2.xlarge" machines and the elastic machines consist of $4$ \verb"t2.large" and $4$ \verb"t2.xlarge" machines for a total of $18$ worker machines.



\subsection{Heterogeneous Computation Speed}

In Table \ref{table: comp_speed}, we give an example of the measured relative computation speeds of  Amazon EC2 instances during a trial of the power iteration and gradient descent algorithm for linear regression, resp.,  using CSEC. The slowest speed is shown in blue and the highest speed is shown in red. 
One interesting observation is that the speeds can vary widely even for the same instance type with the same configurations. 
For example, in the linear regression application, the normalized computation speed of the \verb"t2.large" instances varies from $0.48$ to $0.90$ and the \verb"t2.xlarge" instances vary from $0.58$ to $1.37$. 
Furthermore, while  not shown here, we observed the computation speeds remained nearly constant on the order of minutes and the speeds are predictable between computation steps which are on the order of seconds.

\begin{table}[h!]
\caption{Normalized Computation Speed}
\label{table: comp_speed}
\centering
\begin{tabular}{ |>{\centering}m{0.55cm}>{\centering}m{0.55cm}|>{\centering}m{0.55cm}>{\centering}m{0.55cm}|| >{\centering}m{0.55cm}>{\centering}m{0.55cm}|>{\centering}m{0.55cm}>{\centering}m{0.55cm}| }
\hline
\multicolumn{4}{|c||}{Power Iteration}& \multicolumn{4}{c|}{Linear Regression}\\
\hline
\multicolumn{2}{|c|}{\cs{2.large}}&\multicolumn{2}{c||}{\cs{2.xlarge}}&\multicolumn{2}{c|}{\cs{2.large}}&\multicolumn{2}{c|}{\cs{2.xlarge}} \\
\hline
$0.85$ & $0.84$ & $1.28$ & $1.20$ & $0.70$ & $0.53$ & $1.30$ & $1.26$ \\
$0.76$ & $0.88$ & $0.92$  & $1.27$ & $0.90$ & $0.90$ & $1.26$ & $1.34$ \\
$0.59$ & $0.64$ & $1.20$ & $0.91$  & $0.88$ & $0.77$ & $0.88$  & $1.26$ \\
${\BLUE 0.59}$ & $0.88$ & ${\RED 1.34}$ & ${\BLUE 0.85}$  & ${\RED 0.90}$ & ${\BLUE 0.48}$ & $1.23$ & ${\RED 1.37}$\\
${\RED 0.90}$ & $0.70$ & $1.29$ & $0.90$  & $0.52$ & $0.48$ & $0.83$  & ${\BLUE 0.58}$\\
\hline
\end{tabular}
\end{table}

\subsection{Evaluation Results}

\begin{figure}
\hspace{-0.5cm}
{\centering
\includegraphics[width=1\columnwidth]{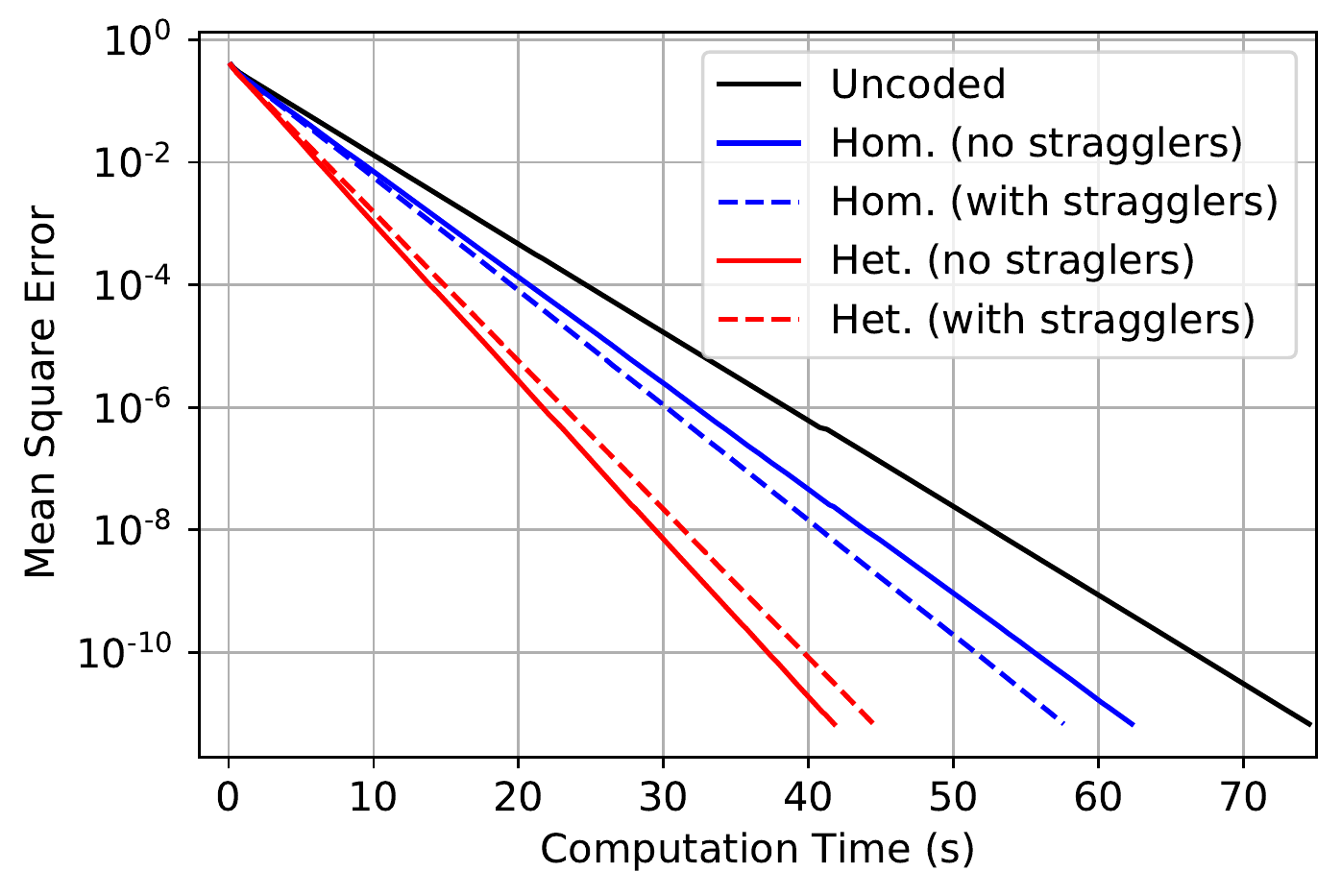}}
\caption{{\bf Power Iteration}: Results using CSEC designs on Amazon EC2 without stragglers and with $2$ stragglers each iteration.  The y-axis represents the normalized mean square error between the true dominant eigenvector and the estimated eigenvector. 
}
\label{fig: Power}
\end{figure}

\begin{figure}
\hspace{-0.5cm}
{\centering
\includegraphics[width=1\columnwidth]{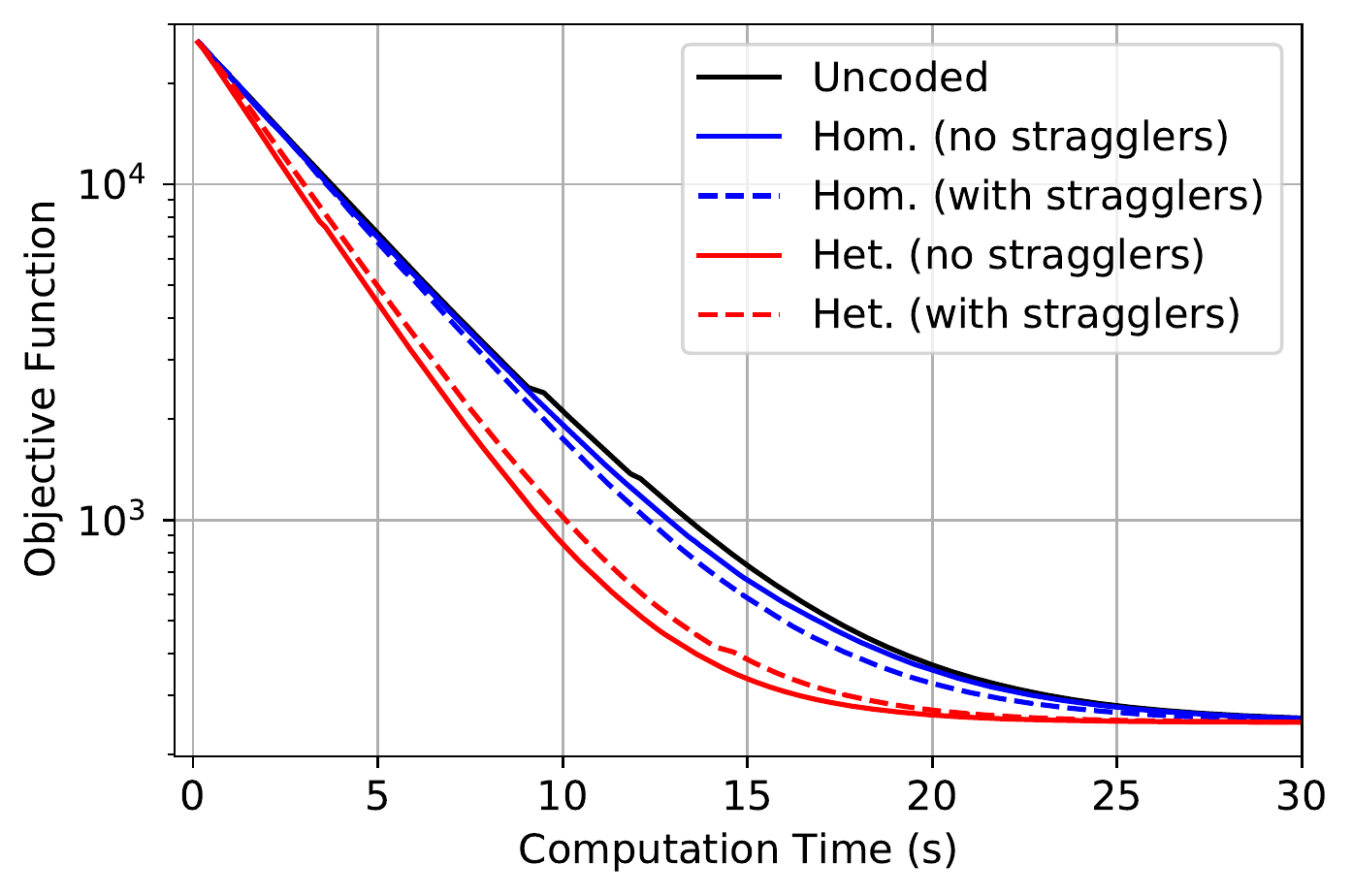}}
\caption{{\bf Linear Regression}: Results using CSEC designs on Amazon EC2 without stragglers and with $2$ stragglers each iteration. 
}
\label{fig: LinRegr}
\vspace{-0.5cm}
\end{figure}

The results of using the CSEC designs on Amazon EC2 are shown in Figs. \ref{fig: Power} and \ref{fig: LinRegr} for  power iteration and linear regression, respectively. We observe similar trends for both applications. 
When no straggler tolerance is considered, the results are shown as solid lines in Figs. \ref{fig: Power} and \ref{fig: LinRegr}. We can see that the proposed heterogeneous CSEC algorithm converges faster than other cases. In particular, for power iteration, nearly 40\% gain can be obtained compared to the uncoded design and about 30\% gain can be obtained compared to the homogeneous design. This occurs because each computation step, which updates the vector or weights, is performed faster when compared to the other designs. Both the uncoded design and homogeneous cyclic design assign the same computation load to each machine. These designs are hindered by slower machines with lower computation speed, even though they may share the same configurations. Faster machines will complete and then wait for slower machines to finish. However, for the proposed heterogeneous design, computation load is assigned relative to computation speed so that ideally each machine will finish at approximately the same time regardless of its speed.

There are several key observations from the experiments with straggler tolerance whose results are shown as  dashed lines in Figs. \ref{fig: Power} and \ref{fig: LinRegr}. 
First, the heterogeneous CSEC design converges faster than the homogeneous CSEC design for similar reasons as those of the experiments without straggler tolerance. 
Second, the straggler tolerance algorithms improve the performance of homogeneous designs. This is because the slowest machines are slow enough so that ignoring them while assigning more tasks to other machines can help. However, interestingly, for the proposed heterogeneous design, since the slowest machines are optimally assigned less tasks and the speed of machines changes slowly in our experiments, the algorithms without straggler tolerance can be better.

\section{Conclusion}
\label{sec: conclusion}
We introduce straggler tolerance in CSEC systems and bridge the gap between CSEC and straggler tolerant distributed computing. We demonstrate a trade-off between computation time and straggler tolerance under our new general framework. We validate our straggler tolerant CSEC designs on Amazon EC2 for the applications of the power iteration and gradient descent for linear regression. We find the proposed heterogeneous CSEC designs to be particularly effective and robust in heterogeneous environments where machine speeds can vary widely even though they share the same configurations.


\bibliographystyle{IEEEbib}
\bibliography{references_d2d}

\end{document}